\documentclass[aps,preprint]{revtex4}
\usepackage{amsfonts}
\usepackage{amssymb}
\usepackage{amsmath}
\usepackage{graphicx}
\usepackage{hyperref}
\usepackage{float}
\usepackage{placeins}
\usepackage[font={footnotesize,it}]{caption}
\usepackage{latexsym}
\usepackage{xcolor}
\usepackage{hyperref}

\usepackage{bm}
\begin{document}

\title{Quantum Probe to the Higher Dimensional Yang-Mills Singularity}

\author{Mert Mangut}
\email{mert.mangut@emu.edu.tr}
\affiliation{AS237, Department of Physics, Eastern Mediterranean
University, 99628, Famagusta, North Cyprus via Mersin 10, Turkey}
\author{\"{O}zay G\"{u}rtu\u{g}}
\email{ozaygurtug@beykoz.edu.tr}
\affiliation{Beykoz University, Faculty of Engineering and Architecture,
34810, Istanbul -Turkey}
\author{Mustafa Halilsoy}
\email{mustafa.halilsoy@emu.edu.tr}
\affiliation{Department of Physics, Eastern Mediterranean
University, 99628, Famagusta, North Cyprus via Mersin 10, Turkey}

\begin{abstract}
We investigate the quantum nature of naked curvature singularities in Einstein-Yang-Mills (EYM) theory using the Horowitz-Marolf (HM) criterion, which  assesses quantum singularities via the evolution of quantum scalar fields. Focusing on timelike singularities in spacetime dimension $ D \geq 5 $, we analyze both pure Yang-Mills and Einstein-Maxwell-Yang-Mills (EMYM) solutions. We then incorporate higher curvature corrections through  Gauss-Bonnet (GB) terms. From positivity requirement the expression under square root that arises in GB may create a secondary singularity that shall be scrutinized carefully. Our analysis reveals that while EYM and EMYM spacetimes remain quantum mechanically singular, the inclusion of GB corrections can, in general, render the singularity quantum mechanically regular for specific values of the mass parameter $m$, which is related to the YM charge $Q$ in $ D = 5 $ space-time dimensions. Contrary, for space-time dimension $D \geq 6$, although the outer (secondary) singularity may be healed quantum mechanically for certain values of the mass parameter $m$, the central singularity remains quantum mechanically singular.

\end{abstract}

\maketitle

\section{\label{sec:level1}Introduction}

An exact solution of the non-Abelian Yang-Mills (YM) field equations in $4D$ flat spacetime was given by Wu and Yang \cite{1}. Since then, the method has come to be known as the Wu-Yang ansatz, notable for its intriguing feature that the internal gauge indices became interwind with the space-time indices in the gauge potential $1-$ form given by

\begin{equation}
A^a_idx^i=-\frac{\epsilon^a_{ij}x^jdx^i}{r^2}, \label{1}
\end{equation}
where $\left\{a,b,... \right\}$ and  $\left\{i,j,... \right\}$  are the internal and spatial indices, respectively. Here the symbol $\epsilon^a_{ij}$ is the Levi-Civita symbol. Being independent of time such an ansatz represents a pure magnetic potential satisfying the flat space YM equations. The non-Abelian YM fields in curved spacetime was considered by Yasskin \cite{2}. Extension of the Wu-Yang ansatz to the curved space was accomplished in \cite{3}  $(D=5)$ and \cite{4}  $(D>5)$. It was realized also that the $4D$ Wu-Yang ansatz in a curved space reduces \cite{5}  to Abelian electromagnetic solution of Reissner-Nordstrom (RN). For $D\geq5$ such a reduction is not possible and the resulting solutions are genuinely non-Abelian. To check these it suffices to calculate the YM invariants $I_1=F_{\mu\nu}^aF^{a\mu\nu}$ and $I_2=F_{\mu\nu}^a\overset{\star}{F}\,{}^{a\mu\nu}$, in which $\overset{\star}{F}$ stands for the dual field. It turns out that for $D=4$ the invariants coincide with those of RN showing that we are still in the Abelian sector. \\

In higher-dimensional $(D \geq 5)$ Einstein-Yang-Mills (EYM) gravity, the intricate interplay between non-Abelian gauge fields and space-time curvature often gives rise to reach geometric structures, including a variety of curvature singularities. Understanding the physical nature of these singularities remains a central challenge, particularly in theories where extra dimensions and gauge interactions fundamentally alter gravitational dynamics. This is the main motivation in this study to consider higher-dimensional EYM solutions. \\

In this work, we employ a quantum scalar field as a diagnostic probe to assess the severity and potential resolution of such singularities. By analysing the behaviour of test fields near regions of divergent curvature, we aim to shed light on the quantum accessibility of these singularities and explore whether quantum effects can distinguish  between physically benign and pathological geometries within higher-dimensional EYM solutions. \\

The higher dimensional static EYM spacetime is given in  the generic form

\begin{equation}
ds^2=-f(r)dt^2+\frac{dr^2}{f(r)}+r^2d\Omega^2_{D-2}, \label{2}
\end{equation}
where $d\Omega^2_{D-2}$ refers to the line element on $S^{D-2}$, $f(r)$ is only a function of $r$ and $5\leq D<\infty$. For the exact solutions the metric function $f(r)$ is given \cite{3,4} as follows

\begin{equation}
f(r)= \begin{cases}
  1-\frac{m}{r^2}-\frac{2Q^2}{r^2}ln(r)& \text{for } D=5,\\
  1-\frac{m}{r^{D-3}}-\frac{(D-3)Q^2}{(D-5)r^2}& \text{for } D>5. \label{3}
\end{cases}
\end{equation}

Our notation is such that $m$ stands for the mass of the source and $Q$ is the only non-vanishing YM charge. For $SO(N)$ although we can have at most $N$ different gauge charges we choose them all equal to $Q$. An interesting property of these solutions for $D>5$, which deserves to record is that the YM term in the metric function has the fixed power of the form $\sim \frac{Q^2}{r^2}$, irrespective of the higher dimensions. As observed, for $D=5$ the metric function $f(r)$ has the distinctive property to contain a $ln(r)$ term which does not arise for $D>5$. We note that the $ln(r)$ term is to be understood as $ln\left(\frac{r}{r_0} \right)$, with the choice of the constant $r_0$ as $1$. It will be seen in the following sections that the occurrence of $ln(r)$ term will give much technical complication while probing the singularity with a scalar field. The common feature in all these metrics for $5\leq D<\infty$, is that they have a spacetime singularity at $r=0$, which is timelike, where the Kretchmann scalar diverges. As discussed before, the solution \eqref{3} admits black hole (BH) solutions, however, our interest in this paper is to concentrate on not BHs but naked timelike singularities.\\

Our aim in this paper is to investigate the $r=0$ singularity from a quantum perspective. To this end we send a scalar probe and apply the method developed by Horowitz and Marolf (HM) \cite{6}. To show the applicability of the HM technique we prove that the naked singularity $r=0$ is a timelike one. Upon this we apply first the HM technique to the pure YM metrics for $D\geq5$. Then we add Maxwell field and consider the similar probe for the Einstein-Maxwell-Yang-Mills (EMYM) solution. For the EMYM theory we have the solution \cite{4}

\begin{equation}
f(r)= \begin{cases}
  1-\frac{m}{r^2}-\frac{2Q^2}{r^2}ln(r)+\frac{4q^2}{3r^4}& \text{for } D=5\\
  1-\frac{m}{r^{D-3}}-\frac{(D-3)Q^2}{(D-5)r^2}+\frac{(D-3)2q^2}{(D-2)r^{2(D-3)}}& \text{for } D>5 \label{EMYM}
\end{cases}
\end{equation}
in which $q$ represents the electric charge. Our final test covers the addition of Gauss-Bonnet (GB) terms to the YM solutions. Then EYMGB solution is summarized by \cite{5i}

\begin{equation}
f(r)= \begin{cases}
  1+\frac{r^2}{4\alpha}\left[1\mp\sqrt{1+\frac{16\alpha^2}{r^4}\left(1+\frac{m}{2\alpha} \right)+\frac{16\alpha Q^2}{r^4}ln(r)} \right]& \text{for } D=5\\
  1+\frac{r^2}{2\bar{\alpha}}\left[1\mp\sqrt{1+\frac{8m\bar{\alpha}^2}{r^{D-1}}+\frac{4\bar{\alpha} Q^2(D-3)}{(D-5)r^4}} \right]& \text{for } D>5 \label{EMGBYM}
\end{cases}
\end{equation}
where $\bar{\alpha}=(D-3)(D-4)\alpha$. We note that existence of the square roots compells us to investigate the probe to the $(+)$ outer root \cite{R1,R2}, i.e. $r_{\star}$ for $D=5$ and $\tilde{r}_{\star}$ for $D>5$, in the expressions under square roots. We summarize our results as follows:\\

$i)$ EYM spacetimes are quantum mechanically singular with respect to a scalar probe irrespective of the dimensions $D\geq5$.\\

$ii)$ Addition of Maxwell fields to the YM field does not remove the singularity under a quantum probe. That is, by a quantum probe the Maxwell field does not heal the classical singularity of EMYM spacetime.\\

$iii)$ The addition of the Gauss-Bonnet (GB) term to the EYM theory for $D = 5$ has been shown to heal the classical singularity from a quantum perspective in accordance with the HM prescription. This healing, however, is not universal; it depends sensitively on the value of the mass parameter $m$, which is itself tied to the YM charge $Q$. In contrast, the situation in space-time dimensions $D \geq 6$ is markedly different. Although the apparent secondary singularity can be smoothed out  quantum mechanically, the central singularity at $r=0$ persist and remains quantum mechanically singular, indicating that the GB term correction alone is insufficient to regularize the geometry in higher-dimensional EYM solutions.

Organization of the paper is as follows. In section \ref{sec:level3}, we briefly review the basis and the necessity for probing the classical YM singularities with quantum wave packets.  In section \ref{sec:level4}, the analysis of the timelike naked singularities developed in three different types of solutions in YM theory, namely,  pure YM, EMYM and EYMGB geometries are rigorously investigated.  We summarize our findings in Conclusion and Discussion which appears in section \ref{sec:level5}.

\section{\label{sec:level3}The Basis of Quantum Probe of Spacetime Singularity}
In classical general relativity, one of the most complex and challenging topics is the understanding of spacetime singularities. Among these, the naked singularity - one that is not concealed by an event horizon - appears to be the most significant, especially in the context of Penrose’s weak cosmic censorship hypothesis \cite{pen}. The formation of a naked singularity disrupts the deterministic dynamics of spacetime, rendering it non-globally hyperbolic, or in simpler terms, singular. In classical general relativity, a spacetime singularity is often considered as the incompleteness of geodesics, particularly when probed by point particles. As a result, curvature invariants become unbounded, and the laws of physics cease to hold, leading to a violation of the theory’s deterministic nature.

It is widely accepted among physicists that a consistent theory of quantum gravity could potentially "resolve" these classical singularities. In the absence of such a theory, several approaches have been suggested to integrate quantum mechanics. A notable contribution in this regard was made by Wald \cite{7}. He proposed that a well-defined, fully deterministic dynamical evolution of a Klein-Gordon scalar field, propagating through an arbitrary static spacetime with a timelike singularity, could be employed to probe the singularity. In this approach, the dynamical evolution is framed as the problem of finding self-adjoint extensions of the spatial part of the wave operator. Building upon this idea, HM further suggested that spacetime is quantum mechanically nonsingular if the evolution of a scalar wave packet in any state is \textit{uniquely} determined for all time. Conversely, spacetime is quantum mechanically singular if the spatial component of the scalar wave operator is not \textit{uniquely} determined for all time.
The massive scalar wave packet propagating according to the Klein-Gordon equation can be expressed by separating the time and spatial components in the following form:

\begin{equation}
\frac{\partial ^{2}\psi }{\partial t^{2}}=VD^{i}\left( VD_{i}\psi\right)-V^{2}\tilde{m}^{2}\psi=-\mathcal{A} \psi, \label{4}
\end{equation}
where $V^{2}=-\xi _{\mu }\xi ^{\mu }$ and $D_{i}$ is the spatial covariant derivative on a spatial hypersurface $\Sigma$, which is orthogonal to the static Killing field $\xi^{\mu}$. The spatial operator $\mathcal{A}$ is defined on the Hilbert space $\mathcal{H}$ of  square integrable functions on spatial slice $\Sigma$ $(\mathcal{L}^{2}(\Sigma))$. \\
Here $\mathcal{A}$ is a positive symmetric operator and its self-adjoint extension is always possible. At this point, it is important to highlight that a consistent quantum theory for a single relativistic particle obeying the Klein-Gordon equation in static, globally hyperbolic spacetime was established by Ashtekar and Magnon \cite{8}. This framework has been demonstrated in HM’s method to also apply to static spacetimes that admit a timelike singularity. To probe the singularity, the positive frequency solution of the massive Klein-Gordon scalar wave equation is employed. A key aspect of the HM method is proving that this extension is unique. The wave function for a free relativistic particle is given by \cite{6},

\begin{equation}
i\frac{d\psi }{dt}=\sqrt{\mathcal{A}_{E}}\psi , \label{5}
\end{equation}
whose solution represents the time evolution of a scalar wave with a prescribed initial condition
\begin{equation}
\psi \left( t\right) =e^{-it\sqrt{\mathcal{A}_{E}}}\psi \left( 0\right). \label{6}
\end{equation}
In this expression $\mathcal{A}_{E}$ denotes the self adjoint extension of the operator given in Eq. \eqref{4}. If the operator $\mathcal{A}_{E}$ has \textit{unique} extension, then it is considered essentially self-adjoint, which implies that the classical singularity is regularized in a quantum mechanical sense.
 \\
The essential self-adjointness of the operator $\mathcal{A}_{E}$ can be checked by using the concept of \textit{deficiency indices} \cite{10}. This is achieved by considering the solutions of the equation

\begin{equation}
\left( \mathcal{A}^{\ast }\pm i\right) \psi =0. \label{7}
\end{equation}
If the solutions are not square integrable for all space, then they do not belong to the Hilbert space \cite{9}, this result leads to state that the spatial wave operator $\mathcal{A}$ is essentially self adjoint (see references, \cite{11,12,13}  for mathematical details).  \\
It is important to emphasize here that the method proposed by HM cannot be applied to probe spacelike singularities. The HM prescription is based on analyzing the essential self-adjointness of the spatial part of the Klein-Gordon operator in a static spacetime, which allows one to interpret the evolution of quantum wave packets through a well-defined time-independent Hamiltonian. This construction implicitly assumes the existence of a global timelike Killing vector field so that the wave equation can be separated into time and spatial parts and the dynamics can be governed by time-independent operator. Such a formulation is appropriate for studying timelike singularities, where the spacetime geometry is static and the singularity persist along the time direction. In contrast, spacelike singularities are inherently time-dependent structures that occur at specific moments in the evolution of spacetime, as in gravitational collapse scenarios. In this case, the metric coefficients depend explicitly on time, preventing the separation of variables required to construct a time-independent Hamiltonian operator. Consequently, the quantum evolution must be described by a time-dependent Hamiltonian operator and the self-adjointness analysis employed in the HM approach is no longer directly applicable. For this reason, the HM method is restricted to probing timelike singularities in static spacetimes. 

For analyzing the timelike character of the singularity, we consider a general static singular surface $\phi=r-r_0$, for $r_0=const.$ To ensure that this surface is timelike, we must check the gradient square of the surface to be space-like. If this is the case, then the surface itself is timelike, and therefore the HM scheme becomes applicable. For this purpose, we check the character of the singular surface with the following calculation, which is written for the generic metric \eqref{2},

\begin{equation}
\left( \nabla\phi\right)^2=g^{rr}\phi_{,r}^2, \label{t1}
\end{equation}
so that in order to have this expression $(-)$, i.e., time-like, we must guaranty that $g^{rr}=f(r)>0$ must be satisfied. This crucial condition  holds only for the following three different cases. The character of the singularity in the other solutions is space-like and hence will not be taken into consideration.

\subsection{Pure EYM Case}

In the case of $D=5$ the generic metric is given by

\begin{equation}
f(r)=1-\frac{m}{r^2}-\frac{2Q^2}{r^2}ln(r) \label{t2}.
\end{equation}

For $r\rightarrow0$ we have the last term as dominant and we see that the $r=0$ is a time-like singularity.

\subsection{ EMYM Case}

For $D=5$ the metric function can be written as

\begin{equation}
f(r)=1-\frac{m}{r^2}-\frac{2Q^2}{r^2}ln(r)+\frac{4q^2}{3r^4} \label{tt2}.
\end{equation}

In the limit of $r\rightarrow0$ we have the Maxwell term as dominant and we observe that the $r=0$ is a time-like singularity.

For the scenario in which $D>5$, the general form of the metric can be expressed as

\begin{equation}
f(r)=1-\frac{m}{r^{D-3}}-\frac{(D-3)Q^2}{(D-5)r^2}+\frac{(D-3)2q^2}{(D-2)r^{2(D-3)}}\label{t3}.
\end{equation}

Since the Maxwell term outweighs the YM term as  $r\rightarrow0$, it follows that  $f(r)>0$. This makes the $r=0$ singularity timelike so that HM probe method is applicable.

\subsection{ EYMGB Case}
Let us first consider the case when the spacetime dimension is $D=5$. Two possible scenarios must be taken into consideration; \\
a) The square root term has a real, positive root, denoted by $r_{\star}$. For this particular case the square root term can be expressed in the following form

\begin{equation}
\sqrt{}=\sqrt{\left(r-r_{\star} \right)g(r)}\label{t4}.
\end{equation}

It is evident that the minimum accessible radial distance is $r=r_{\star}$. Here, the function $g(r)$ is an analytic function  and we have to check the limit when $r\rightarrow r_{\star}$, which implies that

\begin{equation}
f(r)\underset{r\rightarrow r_{\star}}{\rightarrow} 1+\frac{ r_{\star}^2}{4\alpha}>0, \; for \;\; D=5 \;\; \label{t5}.
\end{equation}

This guarantees that the singularity $r=r_{\star}$ is time-like, apt for the HM probes.\\

b) If the square root term does not admit a real root, then as $r\rightarrow0$, the $ln(r)$ term dominates and contributes a negative divergence under the square root. This leads to an unphysical (imaginary) result. To avoid this issue, we choose $\alpha=-\vert \alpha \vert$ in the case of the 'minus' metric choice, along with a negative sign in front of the square root term. \\
On the other hand, for the 'plus' metric choice, consistency requires $\alpha > 0$, i.e.,  $\alpha$ must be positive. \\

As in the previous solution, the existence of time-like singularities in this case can also be analyzed under two distinct scenarios for the spacetime dimension $D>5$. First, consider the case in which the square root term possesses a real, positive root, denoted by $\tilde{r}_{\star}$. As in the previous case, the metric function similarly reduces to the following form at the point $r=\tilde{r}_{\star}$.

\begin{equation}
f(r)\underset{r\rightarrow \tilde{r}_{\star}}{\rightarrow} 1+\frac{ \tilde{r}_{\star}^2}{2\bar{\alpha}}>0, \; for \;\; D>5 \;\; \label{t6}.
\end{equation}

Therefore, the $\tilde{r}_{\star}-$singularity remains time-like in this case, and the HM analysis can be applied. \\
If the square root term lacks a real root, it becomes dominant in the limit $r\rightarrow0$. However, in this regime, the term under the square root becomes negative, leading to an unphysical (imaginary) result. To circumvent this issue, we adopt $\alpha=-\vert \alpha \vert$ for the 'minus' metric choice, along with a negative sign in front of the square root term. Conversely, for the 'plus' metric choice, consistency requires $\alpha>0$.

\section{\label{sec:level4}Scalar Quantum Probe of the Yang-Mills Singularity}
The $D$-dimensional static spherically symmetric line element considered in this study is given by
\begin{equation}
ds^2=-f(r)dt^2+\frac{dr^2}{f(r)}+r^2d\Omega^2_{D-2}, \label{8}
\end{equation}
in which, $d\Omega^2_{D-2}$ is the line element on $S^{D-2}$, which is given in standard spherical form by
\begin{equation}
d\Omega^2_{D-2}=d\theta^{2}_{1}+\sum_{i=2}^{D-3}\prod_{j=1}^{i-1}sin^{2}\theta_jd\theta^2_i\;\;
\end{equation}
where, \hspace{1 cm}  $ 0\leq \theta_{1} \leq \pi$  and   $0\leq \theta_{j} \leq 2\pi $. \\

The massless Klein-Gordon equation is given by

\begin{equation}
\partial _{\mu }\left[ \sqrt{-g}g^{\mu \nu
}\partial _{\nu }\right]  \psi =0. \label{9}
\end{equation}%
The explicit form of this equation by splitting time and spatial part can be written as
\begin{align}
\frac{\partial ^{2}\psi }{\partial t^{2}} &=
-f(r) \left( \frac{(D-2)f(r)}{r} + f'(r) \right) \frac{\partial \psi }{\partial r}-f^{2}(r) \frac{\partial ^{2}\psi }{\partial r^{2}}-\frac{f(r)}{r^{2}} \mathcal{Y}\psi \equiv \mathcal{A}\psi, \label{10}
\end{align}
in which $\mathcal{A}$ represents spatial operator. Here, $\mathcal{Y}$ the angular operator is given by

\begin{equation}
\mathcal{Y}\equiv\frac{1}{sin\theta_1}\frac{\partial}{\partial\theta_1}\left(sin\theta_1 \frac{\partial}{\partial\theta_1}\right)+...+\frac{1}{sin^2\theta_1...sin^2\theta_{D-3}}\frac{\partial^2}{\partial\varphi^2}.  \label{Angu}
\end{equation}

If we separate the variables $\psi = R(r) Y(\theta_1,\theta_{2},...,\theta_j\equiv\varphi)$,  \textcolor{blue}{Eq.\eqref{7}} takes the form

\begin{equation}
R'' + \frac{(r^{D-2} f)'}{f r^{D-2}} R' - \left[ \frac{l}{f r^2}  \pm \frac{i}{f^2} \right] R = 0.  \label{11}
\end{equation}
in which $l$ is the separation constant of angular part and $'$ denotes derivative with respect to $r$.\\
The square integrability of the solutions of Eq.\eqref{11} for each sign $\pm$ is examined by evaluating the squared norm. The function space defined on each $t=constant$ hypersurface $\Sigma_t $ is taken to be the usual Hilbert space  $\mathcal{H}=\{R: \parallel R  \parallel < \infty \}.$ The squared norm for the general $D-dimensional$ spherically symmetric manifold given in Eq. \eqref{8} is expressed as
\begin{equation}
\Vert R\Vert ^{2}\sim \int_{S^{D-2}}f^{-1}r^{D-2}RR^{\ast }dr, \label{14}
\end{equation}
where $\ast$ represents the complex conjugation. If the squared norm of the solutions of Eq.\eqref{11} is finite, the solutions are square integrable and therefore belong to the Hilbert space. In this case, the operator $\mathcal{A}$ is not essentially self-adjoint, since it admits infinitely many self-adjoint extensions. On the other hand, if the squared norm diverges, the solutions are not square integrable and hence do not belong to the Hilbert space. In this situation, the operator $\mathcal{A}$ admits a unique self-adjoint extension and is therefore said to be essentially self-adjoint.

In the next subsections, timelike naked singularities developed in the classical singular spacetimes of EYM, EMYM and Einstein-Gauss Bonnet-YM (EGBYM) theories will be studied with quantum wave packets to see if the singularity is \textit{healed} in view of quantum mechanics.

\subsection{Scalar Quantum Probe of  EYM Solution for $D=5$:}

In this subsection, we will analyze the scalar wave probe for the classical singularity in the EYM solution by using the general radial equation obtained in Eq.\eqref{11} for $D=5$. The wave which will be used for probing the singularity is the solution of Eq. \eqref{11}. It is important to emphasize  that for the sake of obtaining  physically tractable solutions, the analysis in the remainder of this paper focuses solely  on the leading behaviour of the metric function near the singularity $( r \rightarrow 0)$ and in the asymptotic region $( r \rightarrow \infty )$. In this context, the generic behaviours of the metric function are given by

\begin{equation}
 \begin{cases}
f(r) \approx 1-\frac{2Q^2ln(r)}{r^2}& , r\rightarrow 0 \\
f(r) \approx 1& ,  r\rightarrow \infty 
\label{m1}
\end{cases}
\end{equation}

As a result of this limiting cases, Eq.\eqref{11} is separated into two distinct cases as given below.

\begin{equation}
 \begin{cases}
  R^{\prime\prime}+\frac{1}{r}R^{\prime}=0& , r\rightarrow 0 \\
  R^{\prime\prime}+\frac{3}{r}R^{\prime}\pm iR=0& , r\rightarrow \infty \label{12}
\end{cases}
\end{equation}
The solutions of Eq.\eqref{12} can be written as

\begin{equation}
 \begin{cases}
  R(r)=C_1+C_2lnr& , r\rightarrow 0 \\
  R(r)=\frac{C_3}{r}J_1\left(\sqrt{\pm i}r\right)+\frac{C_4}{r}N_1\left(\sqrt{\pm i}r\right)& , r\rightarrow \infty \label{13}
\end{cases}
\end{equation}

in which $J_1$ and $N_1$ are the Bessel and Neumann functions of first order, respectively, while $C_1$, $C_2$, $C_3$, and $C_4$ represent integration constants. Considering the norm formula defined in Eq.\eqref{14} in $D=5$ dimensions and using the general solutions in Eq.\eqref{13}, together with the asymptotic forms $J_{\nu}(z)\sim\sqrt{\frac{2}{\pi z}}\left[\cos(z-\frac{\nu\pi}{2}-\frac{\pi}{4})\right]$ and $N_{\nu}(z)\sim\sqrt{\frac{2}{\pi z}}\left[\sin(z-\frac{\nu\pi}{2}-\frac{\pi}{4})\right]$ as $z\rightarrow\infty$, the norms are obtained as

\begin{equation}
\Vert R\Vert ^{2}\sim \begin{cases}
\int_0^{const} \frac{r^5\left(C_1+C_2lnr\right)^2}{r^2-2Q^2lnr}dr & , r\rightarrow0, \\
 \int_{const}^{\infty} \frac{\left(\left(C_3^2+C_4^2\right) \cosh \left(\sqrt{2} r\right)+(C_3^2-C_4^2) \sin \left(\sqrt{2} r\right)+2 C_3 C_4 \cos \left(\sqrt{2} r\right)\right)}{2} dr& , r\rightarrow \infty.\label{15}
\end{cases}
\end{equation}

The convergence analysis of the listed integrals for two different limit cases in Eq.\eqref{15} can be performed by using the comparative test. As a requirement of the test, the following inequalities are defined for each limiting case,

\begin{equation}
\begin{cases}
\frac{r^5 \left(C_1 + C_2 \ln r \right)^2}{r^2 - 2 Q^2 \ln r}
\underset{r \to 0}{<} 
r^3 \left( (C_1 + C_2) \ln r\ \right)^2, & r \to 0, \\[1ex]
\frac{(C_3^2 + C_4^2) \cosh(\sqrt{2} r) - \left( |C_3^2 - C_4^2| + 2 |C_3 C_4| \right)}{2}
\underset{r \to \infty}{<} \\ 
\frac{ (C_3^2 + C_4^2) \cosh(\sqrt{2} r) + (C_3^2 - C_4^2) \sin(\sqrt{2} r) + 2 C_3 C_4 \cos(\sqrt{2} r) }{2}, & r \to \infty
\end{cases}
\label{m16}
\end{equation}

Since integrals of the form $\int_0^{c} r^a (\ln r)^b \, dr$
converge for any $a > -1$ and finite $b$, it follows that the comparison function $g(r) = r^3 (\ln r)^2$ is integrable at $r = 0$. Hence, by the comparison test and noting that the integrand is positive for sufficiently small $r$, the original integral is convergent in the neighborhood of $r=0$, i.e., $\int_0^{const} \frac{r^5 (C_1 + C_2 \ln r)^2}{r^2 - 2Q^2 \ln r} \, dr < \infty$.
 Moreover, $\int_{const}^{\infty}\frac{\left(C_3^2+C_4^2\right)cosh(\sqrt{2}r)-\left(\vert C_3^2-C_4^2 \vert+2\vert C_3C_4 \vert \right)}{2}dr \rightarrow \infty $, according to the comparison test, $  \int_{const}^{\infty} \frac{\left(\left(C_3^2+C_4^2\right) \cosh \left(\sqrt{2} r\right)+(C_3^2-C_4^2) \sin \left(\sqrt{2} r\right)+2 C_3 C_4 \cos \left(\sqrt{2} r\right)\right)}{2} dr \rightarrow \infty$. \\
The square integrability analysis shows that one of the solutions near $r \rightarrow 0$ is square integrable, whereas the other solution is not. According to the HM criterion, the spatial operator $\mathcal{A}$  must admit a unique self-adjoint extension in order for the spacetime to be considered quantum mechanically regular. This requirement is satisfied only when both independent solutions fail to be square integrable, ensuring that the deficiency indices vanish and the operator is essentially self-adjoint. Since one of the solution is square integrable in the present case, the operator admits multiple self-adjoint extensions. Consequently, the spacetime remains quantum mechanically singular according to the HM criterion. \\

It should be noted that our analysis throughout this paper focuses exclusively on timelike naked singularities. In principle, if multiple self-adjoint extensions exist, one of these extensions may be selected by imposing appropriate boundary conditions through the integration constants appearing in the solution of Eq.\eqref{11}. However, such boundary conditions cannot be introduced arbitrarily; they must be justified by physical requirement associated with the spacetime under consideration. Since our study deals specifically with naked singularities, there is no natural physical criterion available to uniquely determine these boundary conditions. Consequently, we restrict our analysis to cases in which the spatial operator admits a unique self-adjoint extension, without imposing additional conditions on the integration constants. This approach will be adopted throughout the remainder of the paper.

\subsection{Scalar Quantum Probe of EMYM Solution }

In this subsection, we will analyze the scalar wave probe for the classical singularity in the EMYM solution by using the general radial equation obtained in Eq.\eqref{11} for $D=5$ and $D>5$ dimensions, respectively.

\subsubsection{Analysis of wave packets when $D=5$:}
Our focus in this particular case will be the solution obtained for $D=5$. The explicit solution for this case is given in \eqref{EMYM}. The wave that will be used to investigate the singularity is the solution of Eq. \eqref{11}. The behavior of the metric functions near the singularity $( r \rightarrow 0)$ and in the asymptotic region $( r \rightarrow \infty )$ is given by

\begin{equation}
 \begin{cases}
f(r) \approx 1+\frac{4q^2}{3r^4}& , r\rightarrow 0 \\
f(r) \approx 1-\frac{2Q^2ln(r)}{r^2}& , r\rightarrow \infty \label{m1}
\end{cases}
\end{equation}

As a result of this limiting cases, Eq.\eqref{11} is written for each case as

\begin{equation}
 \begin{cases}
  R^{\prime\prime}-\frac{1}{r}R^{\prime}=0& , r\rightarrow 0 \\
  R^{\prime\prime}+\frac{3}{r}R^{\prime}\pm iR=0& , r\rightarrow \infty \label{m12}
\end{cases}
\end{equation}
The solutions of Eq.\eqref{m12} for each particular case

\begin{equation}
 \begin{cases}
  R(r)=C_1+C_2r^2& , r\rightarrow 0 \\
  R(r)=\frac{C_3}{r}J_1\left(\sqrt{\pm i}r\right)+\frac{C_4}{r}N_1\left(\sqrt{\pm i}r\right)& , r\rightarrow \infty \label{m13}
\end{cases}
\end{equation}
in which  $J_1$ and $N_1$ are the Bessel function and the Neumann function of the first order, respectively, while $C_1,$ $C_2,$ $C_3,$ and $C_4$  represent integration constants. \\

In $D=5$ dimension and taking the general solutions in Eq.\eqref{m13}  along with the related metric functions, while accounting for the asymptotic values $J_{\nu}(z)\sim\sqrt{\frac{2}{\pi z}}\left[cos(z-\frac{\nu\pi}{2}-\frac{\pi}{4}) \right]$ and  $N_{\nu}(z)\sim\sqrt{\frac{2}{\pi z}}\left[sin(z-\frac{\nu\pi}{2}-\frac{\pi}{4}) \right]$ at $z\rightarrow\infty$, the norms are obtained

\begin{equation}
\Vert R\Vert ^{2}\sim \begin{cases}
\int_0^{const} \frac{3r^7\left(C_1+C_2r^2\right)^2}{3r^4+4q^2}dr & , r\rightarrow0 \\
 \int_{const}^{\infty} \frac{r^2\left(\left(C_3^2+C_4^2\right) \cosh \left(\sqrt{2} r\right)+(C_3^2-C_4^2) \sin \left(\sqrt{2} r\right)+2 C_3 C_4 \cos \left(\sqrt{2} r\right)\right)}{2\left(r^2-2Q^2lnr \right)} dr& , r\rightarrow \infty \label{m15}
\end{cases}
\end{equation}

The convergence analysis of the listed integrals for two different limit cases in Eq.\eqref{m15} can be performed by using the comparative test. In doing so, the following inequalities are defined for each limiting case,

\begin{equation}
\begin{cases}
\frac{3r^7 \left(C_1 + C_2 r^2 \right)^2}{3r^4 +4q^2} 
\underset{r \to 0}{<} 
\frac{C_1^2}{3r^4 +4q^2}, & r \to 0, \\[1ex]
\frac{
\left(C_3^2+C_4^2\right)cosh(\sqrt{2}r)-\left(\vert C_3^2-C_4^2 \vert+2\vert C_3C_4 \vert \right)}{2 + 4 Q^2} 
\underset{r \to \infty}{<} \\ 
\frac{
r^2 \bigl( (C_3^2 + C_4^2) \cosh(\sqrt{2} r) 
+ (C_3^2 - C_4^2) \sin(\sqrt{2} r) 
+ 2 C_3 C_4 \cos(\sqrt{2} r) \bigr)
}{2 ( r^2 - 2 Q^2 \ln r )}, & r \to \infty
\end{cases}
\label{m16}
\end{equation}

$\int_0^{const} \frac{C_1^2}{3r^4 +4q^2}dr<\infty$ and $ \int_{const}^{\infty}\frac{
\left(C_3^2+C_4^2\right)cosh(\sqrt{2}r)-\left(\vert C_3^2-C_4^2 \vert+2\vert C_3C_4 \vert \right)}{2 + 4 Q^2}dr \rightarrow \infty $, according to the comparison test as in the previous section, the spatial operator $\mathcal{A}$ has no unique extension and therefore the classical singularity remains quantum mechanically singular.

\subsubsection{Analysis of wave packets when $D>5$:}

In the previous case we have analysed the naked singularity in the EMYM theory at $D=5$  with quantum wave packets by considering the leading behaviour of the metric near the singularity and at an asymptotic region. The same method will be used in this subsection to investigate the quantum singularity structure of the EMYM solution given in Eq.\eqref{EMYM}. Our analysis will cover for  $D>5$ dimensions. For the case, the asymptotic behaviors of Eq.\eqref{EMYM} can be written as

\begin{equation}
 \begin{cases}
f(r) \approx 1+\frac{2(n-2)q^2}{(n-1)r^{2(n-2)}}& , r\rightarrow 0 \\
f(r) \approx 1-\frac{(n-2)Q^2}{(n-4)r^2}& , r\rightarrow \infty \label{m4}
\end{cases}
\end{equation}

Note that in the sequel of the paper, it has been found more convenient to replace the dimension parameter $D$ with $D=n+1$. By inserting the asymptotic metric functions into Eq.\eqref{11}, in these two limiting cases and identifying the leading-order terms, Eq.\eqref{11} reduces to

\begin{equation}
 \begin{cases}
  R^{\prime\prime}+\frac{3-n}{r}R^{\prime}=0& , r\rightarrow0 \\
  R^{\prime\prime}+\left(\frac{n-1}{r}\right)R^{\prime}\pm iR=0& , r\rightarrow \infty \label{24}
\end{cases}
\end{equation}
The solutions of Eq.\eqref{24} can be expressed as

\begin{equation}
 \begin{cases}
  R(r)=C_1+C_2r^{n-2}& , r\rightarrow0 \\
  R(r)=r^{\frac{2-n}{2}}J_{\frac{2-n}{2}}\left(\sqrt{\pm i}r\right)C_3+r^{\frac{2-n}{2}}N_{\frac{2-n}{2}}\left(\sqrt{\pm i}r\right)C_4& , r\rightarrow \infty \label{25}
\end{cases}
\end{equation}

 Following the same steps as in the previous analysis, the square norm defined in Eq.\eqref{14} reduces to the following integrals for the two limiting cases

\begin{equation}
\Vert R\Vert ^{2}\sim 
\begin{cases}
\int_0^{const} 
\frac{r^{3n-5}\left(C_1+C_2r^{n-2}\right)^2}
{(n-1)r^{2(n-2)}+2(n-2)q^2}dr
& , r\rightarrow0 \\[1ex]
\int_{const}^{\infty}
\left[\frac{r^2}{2\left(r^2-\frac{(n-2)}{(n-4)}Q^2\right)}\right]\left\{
2 C_4 C_3 \cos\left(\frac{\pi n}{2}-\sqrt{2} r\right)
\right. \\
\left.
+C_3^2\left(\sin\left(\frac{\pi n}{2}-\sqrt{2} r\right)
+ \cosh(\sqrt{2} r)\right)
\right. \\
\left.
+ C_4^2\left(\cosh(\sqrt{2} r)
-\sin\left(\frac{\pi n}{2}-\sqrt{2} r\right)\right)
\right\}dr
& , r\rightarrow\infty
\label{19}
\end{cases}
\end{equation}

The convergence analysis of the given integrals in Eq.\eqref{19} for the two distinct limiting cases can be carried out using a comparative test. As a requirement of this test, the following inequalities are introduced for evaluating the integrals,

\begin{equation}
\begin{cases}
\frac{r^{3n-5}\left(C_1+C_2r^{n-2}\right)^2}{(n-1)r^{2(n-2)}+2(n-2)q^2}
\underset{r\rightarrow0}{<}
\frac{r^{n-1}\left(C_1+C_2r^{n-2}\right)^2}{(n-1)+2(n-2)q^2}
& , r\rightarrow0 \\[1ex]
\frac{\left(C_3^2+C_4^2\right)\cosh(\sqrt{2}r)
-\left(|C_3^2-C_4^2|+2|C_3C_4|\right)}
{2\left(1-\frac{(n-2)}{(n-4)}Q^2\right)}
\underset{r\rightarrow\infty}{<} \\
\frac{r^{2}\left(2 C_4 C_3 \cos \left(\frac{\pi n}{2}-\sqrt{2} r\right)
+ C_3^2 \left(\sin \left(\frac{\pi n}{2}-\sqrt{2} r\right)
+ \cosh \left(\sqrt{2} r\right)\right)
+ C_4^2 \left(\cosh \left(\sqrt{2} r\right)
-\sin \left(\frac{\pi n}{2}-\sqrt{2} r\right)\right)\right)}
{2\left(r^2-\frac{(n-2)}{(n-4)}Q^2\right)}
& , r\rightarrow \infty
\label{20}
\end{cases}
\end{equation}

Since $\int_0^{const}\frac{r^{n-1}\left(C_1+C_2r^{n-2}\right)^2}{(n-1)+2(n-2)q^2}dr<\infty$ and  $ \int_{const}^{\infty}\frac{\left(C_3^2+C_4^2\right)\cosh(\sqrt{2}r)
-\left(|C_3^2-C_4^2|+2|C_3C_4|\right)}
{2\left(1-\frac{(n-2)}{(n-4)}Q^2\right)}dr \rightarrow \infty $, the comparison test implies that  $\int_0^{const}\frac{r^{3n-5}\left(C_1+C_2r^{n-2}\right)^2}{(n-1)r^{2(n-2)}+2(n-2)q^2}dr<\infty$  and  $ \int_{const}^{\infty} \frac{r^{2}\left(2 C_4 C_3 \cos \left(\frac{\pi n}{2}-\sqrt{2} r\right)
+ C_3^2 \left(\sin \left(\frac{\pi n}{2}-\sqrt{2} r\right)
+ \cosh \left(\sqrt{2} r\right)\right)
+ C_4^2 \left(\cosh \left(\sqrt{2} r\right)
-\sin \left(\frac{\pi n}{2}-\sqrt{2} r\right)\right)\right)}
{2\left(r^2-\frac{(n-2)}{(n-4)}Q^2\right)} dr \rightarrow \infty$. \\

In summary, the naked singularity in EMYM theory, irrespective of the spacetime dimension , it is always singular in quantum mechanical framework.

\subsection{Scalar Quantum Probe of EGBYM Solution:}
In the preceding sections, we investigated the naked singularity in the Einstein-Yang-Mills (EYM) and Einstein-Maxwell-Yang-Mills (EMYM) theories using quantum wave packet analysis. This was achieved by studying the leading-order behavior of the metric near the singularity and in the asymptotic regime. In this subsection, we extend the same methodology to explore the quantum nature of the singularity in the Einstein-Gauss-Bonnet-Yang-Mills (EGBYM) solution presented in Eq.\eqref{EMGBYM}. Our analysis will be carried out separately for the cases of $D=5$ and $D>5$ spacetime dimensions. \\
It is worth to emphasizing  that the singularity structure of the EGBYM solution is qualitatively different from that of the EYM and EMYM solutions. As shown by the Kretschmann scalar in the Appendix, besides the central singularity at $r=0$, additional singularities may arise from the conditions $\sigma(r)=0$ for $D=5$ and $\Sigma(r)=0$  for $D>5$. Irrespective of the spacetime dimension, such a singularity-denoted by $r_{*}$-always satisfies  $r_{*}>0$ and is located in the outermost region. Therefore, when examining naked singularities in the EGBYM solution, our analysis will primarily focus on this outermost singular point.
\subsubsection{Analysis of wave packets when $D=5$:}
In the limit $r\rightarrow \infty$, the generic metric \eqref{EMGBYM} reduces to

\begin{equation}
 \begin{cases}
f(r)\approx 1+\delta r^2& , for \;\; + \;\; sign \\
f(r)\approx 1& , for \;\; - \;\; sign \label{M6}
\end{cases}
\end{equation}

where $\delta = \frac{1}{2\alpha}$. For this particular limiting case Eq.\eqref{11} transforms into

\begin{equation}
 \begin{cases}
  R^{\prime\prime}+\frac{5}{r}R^{\prime}=0& , for \;\; + \;\; sign \\
  R^{\prime\prime}+\frac{3}{r}R^{\prime}\pm iR=0& ,for \;\; - \;\; sign\label{27}
\end{cases}
\end{equation}

The solutions of Eq.\eqref{27} are given by

\begin{equation}
 \begin{cases}
 R(r)=C_1+\frac{C_2}{r^4}& , for \;\; + \;\; sign \\
 R(r)=\frac{C_3}{r}J_1\left(\sqrt{\pm i}r\right)+\frac{C_4}{r}N_1\left(\sqrt{\pm i}r\right)& ,for \;\; - \;\; sign\label{28}
\end{cases}
\end{equation}
in which $C_1,$ $C_2,$ $C_3,$ and $C_4$ are integration constants. Substituting the Eq.\eqref{28} into Eq.\eqref{14}, yields

\begin{equation}
\Vert R\Vert ^{2}\sim \begin{cases}
\int_{const}^{\infty} \frac{r^{3}\left(C_1+C_2r^{-4}\right)^2}{1+\delta r^2}dr & ,  for \;\; + \;\; sign \\
 \int_{const}^{\infty} \frac{\left(\left(C_3^2+C_4^2\right) \cosh \left(\sqrt{2} r\right)+(C_3^2-C_4^2) \sin \left(\sqrt{2} r\right)+2 C_3 C_4 \cos \left(\sqrt{2} r\right)\right)}{2} dr& , for \;\; - \;\; sign \label{29}
\end{cases}
\end{equation}

The convergence behavior of the integrals corresponding to the two distinct cases in Eq.\eqref{29} can be examined using the comparison test. In this context, the following inequalities are established for each case

\begin{equation}
\begin{cases}
\frac{r\left(C_1+C_2 r^{-4}\right)^2}{1+\delta}
\underset{r\rightarrow\infty}{<}
\frac{r^{3}\left(C_1+C_2 r^{-4}\right)^2}{1+\delta r^2}
& , \text{for } + \text{ sign} \\[1ex]
\frac{(C_3^2 + C_4^2) \cosh(\sqrt{2} r)
- \left( |C_3^2 - C_4^2| + 2 |C_3 C_4| \right)}{2}
\underset{r\rightarrow\infty}{<} \\
\frac{(C_3^2+C_4^2) \cosh(\sqrt{2} r)
+ (C_3^2-C_4^2) \sin(\sqrt{2} r)
+ 2 C_3 C_4 \cos(\sqrt{2} r)}{2}
& , \text{for } - \text{ sign}
\label{xx}
\end{cases}
\end{equation}

In the first inequality, observing that $ \frac{r\left(C_1+C_2r^{-4}\right)^2}{1+\delta }\underset{r\rightarrow\infty}{<} \frac{r^{3}\left(C_1+C_2r^{-4}\right)^2}{1+\delta r^2}$ and noting that the integral  $\int_{const}^{\infty} \frac{r\left(C_1+C_2r^{-4}\right)^2}{1+\delta }dr$ is divergent, it follows from the comparison test that the integral $\int_{const}^{\infty} \frac{r^{3}\left(C_1+C_2r^{-4}\right)^2}{1+\delta r^2}dr$ also diverges. In the second inequality, $\int_{const}^{\infty}  \frac{(C_3^2 + C_4^2) \cosh(\sqrt{2} r)
- \left( |C_3^2 - C_4^2| + 2 |C_3 C_4| \right)}{2}dr$ is divergent, according to the comparison test, the integral $\int_{const}^{\infty}\frac{(C_3^2+C_4^2) \cosh(\sqrt{2} r)
+ (C_3^2-C_4^2) \sin(\sqrt{2} r)
+ 2 C_3 C_4 \cos(\sqrt{2} r)}{2}dr$  also diverges. As a result, in the limit $r\rightarrow \infty$, the spatial operator $\mathcal{A}$ does not belong to the Hilbert space. \\

In the limit $r\rightarrow r_{\star}$, analyzing Eq.\eqref{11} in terms of the variable $r$ does not eliminate the nonlinearity. Therefore, we introduce a new coordinate transformation $x=r-r_{\star}$. In the new coordinate, when $r\rightarrow r_{\star}$ $(x<<1)$ the metric function \eqref{EMGBYM}   becomes

  \begin{equation}
f_{\mp}(x)\approx A+Bx \label{30}
\end{equation}

where

\begin{equation}
\begin{aligned}
&A=1+\frac{r_{\star}^2}{4\alpha}\mp\sqrt{1+\frac{m}{2\alpha}+\frac{r_{\star}^4}{16\alpha^2}+\frac{Q^2lnr_{\star}}{\alpha}}\\
&B=\frac{2r_{\star}}{4\alpha}\mp\frac{\frac{r_{\star}^3}{4\alpha^2}+\frac{Q^2}{\alpha r_{\star}}}{2\sqrt{1+\frac{m}{2\alpha}+\frac{r_{\star}^4}{16\alpha^2}+\frac{Q^2lnr_{\star}}{\alpha}}}
\end{aligned}
\end{equation}

When we substitute Eq.\eqref{30} into Eq.\eqref{11}, Eq.\eqref{11}  transforms to

\begin{equation}
\ddot{R}+\left[c-c^2x \right]\dot{R}-\left[b-a^2x \right]R=0 \label{31}
\end{equation}
in which the dot $(\cdot)$ denotes differentiation with respect to $x$. Here, $c=\frac{3}{r_{\star}}+\frac{B}{A}$, $b=\frac{l}{Ar_{\star}}\mp \frac{i}{A^2}$ and $a=\frac{-l}{A^2r_{\star}}\left(\frac{3A}{r_{\star}}+B \right)\pm \frac{2iB}{A^3}$. The solution of Eq.\eqref{31}  can be written as

\begin{equation}
\begin{aligned}
R(x)=&c_1 e^{\frac{a x}{c^2}} H_{\frac{-b c^4+a c^3+a^2}{c^6}}\left(\frac{c x}{\sqrt{2}}-\frac{c^3+2 a}{\sqrt{2} c^3}\right)\\
&+c_2 e^{\frac{a x}{c^2}} \, _1F_1\left(-\frac{-b c^4+a c^3+a^2}{2 c^6};\frac{1}{2};\left(\frac{c x}{\sqrt{2}}-\frac{c^3+2 a}{\sqrt{2} c^3}\right)^2\right) \label{32}
\end{aligned}
\end{equation}
where, $H$ and $\, _1F_1$ represent the Hermite polynomial  and  Kummer confluent hypergeometric function, respectively. Here, $c_1$ and $c_2$ are integration constants. When we substitute Eq.\eqref{32} into Eq.\eqref{14} and using $x<<1$ limiting case, Eq.\eqref{14} becomes

\begin{equation}
\begin{aligned}
\Vert R\Vert ^{2}&\sim \int_0^{const}\frac{C_2r_{\star}^3}{A}\left[1+\left(2C_1e^{\frac{a x}{c^2}} +\frac{3}{r_{\star}}-\frac{B}{A} \right)x \right]dx\\
&\sim\frac{C_2 r_{\star}^2 \left(A e^{\frac{a x}{c^2}} \left(\frac{2 c^2 C_1 r_{\star} x}{a}-\frac{2 c^4 C_1 r_{\star}}{a^2}\right)+A r_{\star} x+\frac{3 A x^2}{2}-\frac{1}{2} B r_{\star} x^2\right)}{A^2} \vert_0^{const}, \label{33}
\end{aligned}
\end{equation}
in which $C_1=2acc_1H_{a-1}(b)+2ac_1c_2\, _1F_1(1+a,3/2,b)$ and $C_2=c_1H_a(b)+c_2\, _1F_1(a,1/2,b)$. If $A\neq0$, the integral \eqref{33} is finite. Thus, in the limit as $r \to r_{\star}$, the spatial operator $\mathcal{A}$ is not essentially self-adjoint. However, in the case of, $A=0$, the integral \eqref{33} becomes divergent. To sum up, under the condition $A=0$, which is valid only when $M= \pm\left(r_{\star}^2-2Q^2ln r_{\star}\right) $, the spatial operator $\mathcal{A}$ does not belong to the Hilbert space and thus the classical singularity becomes quantum mechanically regular. \\

In the case where $\sigma(r) \neq 0$ for $D=5$, the only remaining singularity in the EGBYM solution is the central singularity at $r = 0$, which must then be analyzed within a quantum mechanical framework. In the limit $r \rightarrow 0$, analyzing Eq.\eqref{11} in terms of the variable $r$ does not eliminate the nonlinearity. Therefore, we introduce a new coordinate transformation $e^{-x}=r$. In these new coordinates, Eq.\eqref{11} transforms to

 \begin{equation}
\ddot{R}-2\dot{R}=0, \label{34}
\end{equation}
in which the dot $(\cdot)$ denotes differentiation with respect to $x$ and if $r\rightarrow 0$, then $x\rightarrow \infty$. The solution of Eq.\eqref{34} in $r$ coordinates is given as

\begin{equation}
R(r)=C_3+\frac{C_4}{r^2}. \label{35}
\end{equation}

Here, $C_3$ and $C_4$ represent integration constants. In the limiting case, Eq.\eqref{EMGBYM} becomes

 \begin{equation}
f(r)\approx  1+\bar{\delta} ln^{1/2}r,     \label{emgbym1}
\end{equation}
where $\bar{\delta}=\mp Q/\alpha^{1/2}$. Substituting Eq.\eqref{35} into Eq.\eqref{14}, we obtain

\begin{equation}
\Vert R\Vert ^{2}\sim \int_0^{const} \frac{r^{3}\left(C_3+C_4r^{-2}\right)^2}{1+\bar{\delta} ln^{1/2}r}dr. \label{36}
\end{equation}

The convergence analysis of the integral in Eq.\eqref{36} is carried out using the comparison test, similar to the previous case. From this standpoint, observing that $ \frac{ln(r)\left(C_3+C_4r^{-1}\right)^2}{1+\bar{\delta}}\underset{r\rightarrow0}{<} \frac{r^{3}\left(C_3+C_4r^{-2}\right)^2}{1+\bar{\delta} ln^{1/2}r}$ and considering that the integral  $\int_0^{\text{const}} \frac{ln(r)\left(C_3+C_4r^{-1}\right)^2}{1+\bar{\delta}}dr=\frac{-2 C_3 r+2 C_3 r \log (r)+C_4 \log ^2(r)}{2 (1+\bar{\delta})}\vert_0^{const}$ is divergent, it follows from the comparison test that $\int_0^{\text{const}}\frac{r^{3}\left(C_3+C_4r^{-2}\right)^2}{1+\bar{\delta} ln^{1/2}r}dr \rightarrow\infty.$ Thus, in the limit as $r \to 0$, the solution does not belong to the Hilbert space. \\
In conclusion, the spatial operator $A$ is found to be essentially self-adjoint, ensuring that the evolution of quantum scalar fields is uniquely defined throughout the entire spacetime, from the singular point to infinity. According to the HM criterion, the timelike naked singularities in the EGBYM solution is therefore rendered quantum mechanically regular in the $D=5$ spacetime dimension.

\subsubsection{Analysis of wave packets when $D>5$:}

In the limit $r\rightarrow \infty$, Eq.\eqref{EMGBYM} becomes

\begin{equation}
 \begin{cases}
f(r)\approx 1+\frac{r^2}{\bar{\alpha}}& , for \;\; + \;\; sign \\
f(r)\approx 1& , for \;\; - \;\; sign \label{MM6}
\end{cases}
\end{equation}

By inserting the asymptotic metric functions \eqref{MM6} into Eq.\eqref{11}, Eq.\eqref{11} reduces to

\begin{equation}
 \begin{cases}
R^{\prime\prime}+\frac{(n+1)}{r}R^{\prime}=0& , for \;\; + \;\; sign \\
R^{\prime\prime}+\left(\frac{n-1}{r}\right)R^{\prime}\pm iR= 0& ,for \;\; - \;\; sign\label{37}
\end{cases}
\end{equation}

The solutions of Eq.\eqref{37} can be written as

\begin{equation}
 \begin{cases}
 R(r)=C_1+\frac{C_2}{r^{n-4}}& , for \;\; + \;\; sign \\
 R(r)=r^{\frac{2-n}{2}}J_{\frac{2-n}{2}}\left(\sqrt{\pm i}r\right)C_3+r^{\frac{2-n}{2}}N_{\frac{2-n}{2}}\left(\sqrt{\pm i}r\right)C_4& ,for \;\; - \;\; sign\label{38}
\end{cases}
\end{equation}
in which $C_1,$ $C_2,$ $C_3,$ and $C_4$ represent integration constants. For the solutions \eqref{38}, Eq.\eqref{14} yields

\begin{equation}
\Vert R\Vert ^{2}\sim 
\begin{cases}
\int_{const}^{\infty}
\frac{r^{n-1}\left(C_1+C_2r^{4-n}\right)^2}
{1+\frac{r^2}{\bar{\alpha}}}dr
& , for \;\; + \;\; sign \\[1ex]
\int_{const}^{\infty}
\left(\frac{1}{2}\right)
\left\{
2 C_4 C_3 \cos\left(\frac{\pi n}{2}-\sqrt{2} r\right)
\right. \\
\left.
+ C_3^2\left(\sin\left(\frac{\pi n}{2}-\sqrt{2} r\right)
+ \cosh(\sqrt{2} r)\right)
\right. \\
\left.
+ C_4^2\left(\cosh(\sqrt{2} r)
-\sin\left(\frac{\pi n}{2}-\sqrt{2} r\right)\right)
\right\} dr
& , for \;\; - \;\; sign
\label{39}
\end{cases}
\end{equation}

As demonstrated in the previous sections, the comparison test has proven to be an effective method for evaluating integrals of the form given in Eq.\eqref{39}. Accordingly, we will apply the comparison test to evaluate the integral in Eq.\eqref{39}. To proceed, we establish an appropriate inequality, which will serve as the basis for the calculation. For the first integral, the corresponding inequality is $\frac{r^{n-3}\left(C_1+C_2r^{4-n}\right)^2}{1+\frac{1}{\bar{\alpha}}}\underset{r\rightarrow\infty}{<} \frac{r^{n-1}\left(C_1+C_2r^{4-n}\right)^2}{1+\frac{r^2}{\bar{\alpha}}}$, and  the calculation of the integral  $\int_{const}^{\infty} \frac{r^{n-3}\left(C_1+C_2r^{4-n}\right)^2}{1+\frac{1}{\bar{\alpha}}}dr$, yields a divergent result. The second integral is also divergent because the relevant inequality, $\frac{\left(C_3^2+C_4^2\right)\cosh(\sqrt{2}r)
-\left(|C_3^2-C_4^2|+2|C_3C_4|\right)}
{2}\underset{r\rightarrow\infty}{<}\frac{2 C_4 C_3 \cos \left(\frac{\pi  n}{2}-\sqrt{2} r\right)+C_3^2 \left(\sin \left(\frac{\pi  n}{2}-\sqrt{2} r\right)+\cosh \left(\sqrt{2} r\right)\right)+C_4^2 \left(\cosh \left(\sqrt{2} r\right)-\sin \left(\frac{\pi  n}{2}-\sqrt{2} r\right)\right)}{2}$, in accordance with the comparison test, indicates that the corresponding integral $\int_{const}^{\infty}\frac{\left(C_3^2+C_4^2\right)\cosh(\sqrt{2}r)
-\left(|C_3^2-C_4^2|+2|C_3C_4|\right)}
{2}dr$, yields also a divergent result. \\

Next, we investigate the outermost singularity that may arise from the condition $\Sigma(r)=0$, which we denote by $\tilde{r}_{\star}$. In the limit $r\rightarrow \tilde{r}_{\star}$, the generic metric \eqref{EMGBYM} behaves as

  \begin{equation}
\tilde{f}_{\mp}(\tilde{x})\approx \tilde{A}+\tilde{B}\tilde{x} \label{40}
\end{equation}
in which $\tilde{x}=r-\tilde{r}_{\star}$. Here, the coefficients are given by

\begin{equation}
\begin{aligned}
&\tilde{A}=1+\frac{\tilde{r}_{\star}^2}{2\bar{\alpha}}\left[1\mp\sqrt{1-8m\bar{\alpha}\tilde{r}_{\star}^{1-D}+\frac{4\bar{\alpha}Q^2(D-3)}{(D-5)\tilde{r}_{\star}^4}} \right]\\
&\tilde{B}=\frac{\tilde{r}_{\star}}{\bar{\alpha}}\left[1\mp\sqrt{1-8m\bar{\alpha}\tilde{r}_{\star}^{1-D}+\frac{4\bar{\alpha}Q^2(D-3)}{(D-5)\tilde{r}_{\star}^4}} \right]+\frac{8m\bar{\alpha}(1-D)\tilde{r}_{\star}^{-D}+\frac{16\bar{\alpha}Q^2(D-3)}{(D-5)\tilde{r}_{\star}^5}}{4\bar{\alpha}\sqrt{1-8m\bar{\alpha}\tilde{r}_{\star}^{1-D}+\frac{4\bar{\alpha}Q^2(D-3)}{(D-5)\tilde{r}_{\star}^4}}} \label{41}
\end{aligned}
\end{equation}

By substituting  Eq.\eqref{40} into Eq.\eqref{11}, Eq.\eqref{11} gives

 \begin{equation}
\ddot{R}+\left[\tilde{c}-\tilde{c}^2\tilde{x} \right]\dot{R}-\left[\tilde{b}-\tilde{a}^2\tilde{x} \right]R=0 \label{42}
\end{equation}
in which the dot $(\cdot)$ denotes differentiation with respect to $\tilde{x}$. Here, $\tilde{c}=\frac{n-1}{\tilde{r}_{\star}}+\frac{\tilde{B}}{\tilde{A}}$, $\tilde{b}=\frac{l}{\tilde{A}\tilde{r}_{\star}}\mp \frac{i}{\tilde{A}^2}$ and $\tilde{a}=\frac{-l}{\tilde{A}^2\tilde{r}_{\star}}\left(\frac{3\tilde{A}}{\tilde{r}_{\star}}+\tilde{B} \right)\pm \frac{2i\tilde{B}}{\tilde{A}^3}$. The solution of Eq.\eqref{42}  is given by

\begin{equation}
\begin{aligned}
R(\tilde{x})=&c_2 e^{\frac{\tilde{a} \tilde{x}}{\tilde{c}^2}} H_{\frac{-\tilde{b} \tilde{c}^4+\tilde{a} \tilde{c}^3+\tilde{a}^2}{\tilde{c}^6}}\left(\frac{\tilde{c} \tilde{x}}{\sqrt{2}}-\frac{\tilde{c}^3+2 \tilde{a}}{\sqrt{2} \tilde{c}^3}\right)\\
&+c_3 e^{\frac{\tilde{a} \tilde{x}}{\tilde{c}^2}} \, _1F_1\left(-\frac{-\tilde{b} \tilde{c}^4+\tilde{a} \tilde{c}^3+\tilde{a}^2}{2 \tilde{c}^6};\frac{1}{2};\left(\frac{\tilde{c} \tilde{x}}{\sqrt{2}}-\frac{\tilde{c}^3+2 \tilde{a}}{\sqrt{2} \tilde{c}^3}\right)^2\right) \label{43}
\end{aligned}
\end{equation}
in which $c_2$ and $c_3$ represents the integration constants. If we put Eq.\eqref{43} into Eq.\eqref{14} and considering the limiting case where $\tilde{x}<<1$ when $r\rightarrow  \tilde{r}_{\star}$, Eq.\eqref{14} reduces to

\begin{equation}
\begin{aligned}
\Vert R\Vert ^{2}&\sim \int_0^{const}\frac{C_2\tilde{r}_{\star}^{n-1}}{\tilde{A}}\left[1+\left(2C_1e^{\frac{\tilde{a} \tilde{x}}{\tilde{c}^2}} +\frac{3}{\tilde{r}_{\star}}-\frac{\tilde{B}}{\tilde{A}} \right)\tilde{x} \right]d\tilde{x}\\
&\sim\frac{C_2 \tilde{r}_{\star}^{n-2} \left(\tilde{A} e^{\frac{\tilde{a} \tilde{x}}{\tilde{c}^2}} \left(\frac{2 \tilde{c}^2 C_1 \tilde{r}_{\star} \tilde{x}}{\tilde{a}}-\frac{2 \tilde{c}^4 C_1 \tilde{r}_{\star}}{\tilde{a}^2}\right)+\tilde{A} \tilde{r}_{\star} \tilde{x}+\frac{3 \tilde{A} \tilde{x}^2}{2}-\frac{1}{2} \tilde{B} \tilde{r}_{\star} \tilde{x}^2\right)}{\tilde{A}^2} \vert_0^{const}, \label{44}
\end{aligned}
\end{equation}
where $C_1=2\tilde{a}\tilde{c}c_2H_{\tilde{a}-1}(\tilde{b})+2\tilde{a}c_2c_3\, _1F_1(1+\tilde{a},3/2,\tilde{b})$ and $C_2=c_2H_{\tilde{a}}(\tilde{b})+c_3\, _1F_1(\tilde{a},1/2,\tilde{b})$. This result reveals that as long as $\tilde{A}\neq 0$, the integral converges and hence the solution belongs to the Hilbert space, indicating that the spatial wave operator $\mathcal{A}$ is not essentially self-adjoint, and the spacetime remains quantum singular.   \\
However, similar to the case of $D=5$ spacetime dimension, the condition $\tilde{A}=0$  constrains the mass parameter to
$\left(m = \frac{1}{2}\,\tilde{r}_{\star}^{\,D-5} \left[ \frac{(D-3)}{(D-5)}\,Q^{2} - \bar{\alpha} - \tilde{r}_{\star}^{2} \right]\right)$, which also leads to a diverging result for $D>5$ spacetime dimensions. Under this condition, the spatial operator $\mathcal{A}$ becomes essentially self-adjoint, ensuring a unique time evolution for quantum scalar fields. \\
Consequently, the apparent outermost singularity becomes quantum mechanically regular. However, the central singularity at $r=0$ needs to be checked for a generic conclusion about the quantum nature of the singularities in EGBYM solutions. \\
To this end, we consider the nonvanishing case of the term $\Sigma(r)$ in the Kretschmann scalar. In this particular case the only singularity to be probed by quantum scalar fields is the central singularity located at $r=0$.  Let us now analyze the nature of the singularity at $r\rightarrow0$.  In this limiting case, Eq.\eqref{EMGBYM} becomes

 \begin{equation}
f(r)\approx  1\mp \sqrt{\frac{2M}{\bar{\alpha}}}r^{\frac{4-n}{2}}.     \label{emgbym2}
\end{equation}

Substituting Eq.\eqref{emgbym2} into the Eq.\eqref{11}, Eq.\eqref{11} gives

\begin{equation}
 \begin{cases}
 R^{\prime\prime}+\frac{7}{2r}R^{\prime}+ar^{-3/2}R=0& , D=6 \\
 R^{\prime\prime}+\frac{4}{r}R^{\prime}+ar^{-1}R=0=0& , D=7\\
 R^{\prime\prime}+\frac{9}{2r}R^{\prime}+ar^{-1/2}R=0 & , D=8\\
 R^{\prime\prime}+\frac{5}{r}R^{\prime}+aR=0 & , D=9\\
 R^{\prime\prime}+\frac{D+1}{2r}R^{\prime}=0 & , D\geq10 \label{45}
\end{cases}
\end{equation}

in which $a=\sqrt{\frac{\bar{\alpha}l^2}{2M}}$. The general solutions to the differential equations listed in Eq.\eqref{45} are given by

\begin{equation}
 \begin{cases}
R(r)=\frac{C_1}{r^{5/4}}N_5\left(4\sqrt{a}r^{1/4} \right)+C_2\,_0F_1\left(6,-4a\sqrt{r} \right)& , D=6 \\
R(r)=\frac{C_3}{r^{3/2}}N_5\left(2\sqrt{a}r^{1/2} \right)+C_4\,_0F_1\left(4,-ar \right)& , D=7\\
R(r)=\frac{C_5}{r^{7/4}}J_{7/3}\left(\frac{4}{3}\sqrt{a}r^{3/4} \right)+\frac{C_6}{r^{7/4}}N_{7/3}\left(\frac{4}{3}\sqrt{a}r^{3/4} \right)& , D=8\\
R(r)=\frac{C_7}{r^{2}}J_{2}\left(\sqrt{a}r \right)+\frac{C_8}{r^{2}}N_{2}\left(\sqrt{a}r \right) & , D=9\\
R(r)=C_9+C_{10}r^{-(D-1)/2}& , D\geq10 \label{46}
\end{cases}
\end{equation}

Here, $C_1,...,C_{10}$ are integration constants. If we consider the approximate values $J_{\nu}\sim\frac{1}{\Gamma(\nu+1)}\left(\frac{x}{2} \right)^{\nu}$, $N_{\nu}\sim-\frac{\Gamma(\nu}{\pi}\left(\frac{2}{x} \right)^{\nu}$ and $\,_0F_1\left(s,0 \right)=1$ for $\nu\neq0$, $s=const.$ and in the limit of $r<<1$, the general solutions of the differential equations given in Eq.\eqref{46} reduce to

\begin{equation}
 \begin{cases}
R(r)\sim C_2+\alpha_1r^{-5/2}& , D=6 \\
R(r)\sim C_4+\alpha_2r^{-3}& , D=7\\
R(r)\sim \alpha_3+\alpha_4r^{-7/2}& , D=8\\
R(r)\sim\alpha_5+\alpha_6r^{-4} & , D=9\\
R(r)\sim C_9+C_{10}r^{-(D-1))/2}& , D\geq10 \label{46}
\end{cases}
\end{equation}

where $\alpha_1=\frac{-C_1\Gamma(5)}{\pi}\left(2\sqrt{a} \right)^{-5}$, $\alpha_2=\frac{-C_3\Gamma(5)}{\pi}\left(\sqrt{a} \right)^{-5}$, $\alpha_3=\frac{C_5}{\Gamma(-4/3)}\left( \frac{2}{3}\sqrt{a}\right)$, $\alpha_4=\frac{-C_6\Gamma(7/3)}{\pi}\left(\frac{3}{2\sqrt{a}} \right)^2$, $\alpha_5=\frac{C_7}{\Gamma(3)}\left(\frac{\sqrt{a}}{2} \right)$ and $\alpha_6=\frac{-C_8\Gamma(2)}{\pi}\left(\frac{2}{\sqrt{a}} \right)^2$. When we put the solutions listed in Eq.\eqref{46} into the generic norm formula  Eq.\eqref{11}, the norms become

\begin{equation}
\Vert R\Vert ^{2}\sim \begin{cases}
\int_0^{const} \frac{\left(C_2+\alpha_1r^{-5/2}\right)^2}{1\mp\sqrt{\frac{2M}{\bar{\alpha}}}r^{-1/2}}r^4dr& , D=6 \\
\int_0^{const}\frac{\left( C_4+\alpha_2r^{-3}\right)^2}{1\mp\sqrt{\frac{2M}{\bar{\alpha}}}r^{-1}}r^5dr& , D=7\\
\int_0^{const}\frac{\left(\alpha_3+\alpha_4 r^{-7/2}\right)^2}{1\mp\sqrt{\frac{2M}{\bar{\alpha}}}r^{-3/2}}r^6dr& , D=8\\
\int_0^{const}\frac{\left(\alpha_5+\alpha_6r^{-4}\right)^2}{1\mp\sqrt{\frac{2M}{\bar{\alpha}}}r^{-2}} r^7dr& , D=9\\
\int_0^{const} \frac{\left(C_9+C_{10}r^{-(D-1))/2}\right)^2}{1\mp\sqrt{\frac{2M}{\bar{\alpha}}}r^{4-(D-1)/2}}r^{n-1}dr& , D\geq10 \label{47}
\end{cases}
\end{equation}

As in the previous sections, we can demonstrate the divergence of the integrals by constructing the inequalities below.

\begin{equation}
 \begin{cases}
\frac{\left(C_2+\alpha_1r^{-5/2}\right)^2}{1\mp\sqrt{\frac{2M}{\bar{\alpha}}}r^{-1/2}}r^4\underset{r\rightarrow0}{<}\frac{\left(C_2+\alpha_1\right)^2}{r^{1/2}\mp\sqrt{\frac{2M}{\bar{\alpha}}}}r^{-1/2}& , D=6 \\
\frac{\left( C_4+\alpha_2r^{-3}\right)^2}{1\mp\sqrt{\frac{2M}{\bar{\alpha}}}r^{-1}}r^5\underset{r\rightarrow0}{<}\frac{\left( C_4+\alpha_2\right)^2}{r\mp\sqrt{\frac{2M}{\bar{\alpha}}}}& , D=7\\
\frac{\left(\alpha_3+\alpha_4 r^{-7/2}\right)^2}{1\mp\sqrt{\frac{2M}{\bar{\alpha}}}r^{-3/2}}r^6\underset{r\rightarrow0}{<}\frac{\left( \alpha_3+\alpha_4\right)^2}{r^{3/2}\mp\sqrt{\frac{2M}{\bar{\alpha}}}}r^{1/2}& , D=8\\
\frac{\left(\alpha_5+\alpha_6r^{-4}\right)^2}{1\mp\sqrt{\frac{2M}{\bar{\alpha}}}r^{-2}} r^7\underset{r\rightarrow0}{<}\frac{\left( \alpha_5+\alpha_6\right)^2}{r^2\mp\sqrt{\frac{2M}{\bar{\alpha}}}}r& , D=9\\
\frac{\left(C_9+C_{10}r^{-(D-1))/2}\right)^2}{1\mp\sqrt{\frac{2M}{\bar{\alpha}}}r^{4-(D-1))/2}}r^{D-2}\underset{r\rightarrow0}{<}\frac{\left(C_9+C_{10}\right)^2}{r^{D-1}\mp\sqrt{\frac{2M}{\bar{\alpha}}}}r^{D-2}& , D\geq10 \label{48}
\end{cases}
\end{equation}

The integration of new integrants give

\begin{equation}
\Vert R\Vert ^{2}\sim 
\begin{cases}
\int_0^{const} \frac{(C_2+\alpha_1)^2}{r^{1/2}\mp\sqrt{\frac{2M}{\bar{\alpha}}}} r^{-1/2} dr
= 2 (C_2+\alpha_1)^2 \ln\Big(\sqrt{r}\mp\sqrt{\frac{2M}{\bar{\alpha}}}\Big)\Big|_0^{const}, & D=6,\\[1mm]
\int_0^{const} \frac{(C_4+\alpha_2)^2}{r\mp\sqrt{\frac{2M}{\bar{\alpha}}}} dr
= (C_4+\alpha_2)^2 \ln\Big(r\mp\sqrt{\frac{2M}{\bar{\alpha}}}\Big)\Big|_0^{const}, & D=7,\\[1mm]
\int_0^{const} \frac{(\alpha_3+\alpha_4)^2}{r^{3/2}\mp\sqrt{\frac{2M}{\bar{\alpha}}}} r^{1/2} dr
= \frac{2}{3} (\alpha_3+\alpha_4)^2 \ln\Big(r^{3/2}\mp\sqrt{\frac{2M}{\bar{\alpha}}}\Big)\Big|_0^{const}, & D=8,\\[1mm]
\int_0^{const} \frac{(\alpha_5+\alpha_6)^2}{r^2\mp\sqrt{\frac{2M}{\bar{\alpha}}}} r dr
= \frac{1}{2} (\alpha_5+\alpha_6)^2 \ln\Big(r^2\mp\sqrt{\frac{2M}{\bar{\alpha}}}\Big)\Big|_0^{const}, & D=9,\\[1mm]
\int_0^{const} \frac{(C_9+C_{10})^2}{r^{D-1}\mp\sqrt{\frac{2M}{\bar{\alpha}}}} r^{D-2} dr
= \frac{1}{D-1} (C_9+C_{10})^2 \ln(D-1) \Big|_0^{const} \\
\quad + \frac{1}{D-1} (C_9+C_{10})^2 \ln\Big(r^{D-1}\mp\sqrt{\frac{2M}{\bar{\alpha}}}\Big) \Big|_0^{const}. & D \ge 10
\end{cases}
\end{equation}

Since all the integrals appearing on the left-hand side of the inequalities converge, it follows from the comparison test that the corresponding integrals converge in all relevant norms. This result implies that the spatial operator $\mathcal{A}$ is not essentially self-adjoint, and therefore the classical timelike naked singularities in spacetime dimension $D\geq6$ are, in general, quantum mechanically singular.

\section{\label{sec:level5}Conclusion and discussion}
The occurrence of space–time singularities in general relativistic solutions remains one of the greatest challenges in theoretical physics, as any physically acceptable description of nature should ultimately be free from such pathologies. Over the decades, considerable effort has been devoted to understanding these singularities and exploring possible mechanisms for their resolution. Despite this progress, the problem continues to be highly nontrivial. \\

In this work, we have investigated the time-like singularities of Yang–Mills (YM) space–times in dimensions  $(D \geq 5)$ using the Horowitz–Marolf (HM) prescription, which employs  quantum probe to diagnose the fate of  classical singularities. In particular, we examined whether the wave function representing a quantum particle evolves smoothly and remains unaffected by the presence of the classical singularity, as required for quantum regularity. \\

Our analysis reveals that both the pure YM space–time and the Maxwell field coupled to a YM background remain quantum mechanically singular. In spacetime dimension $D = 5$, however, the inclusion of the Gauss–Bonnet (GB) term in the EYM theory significantly modifies the near-origin geometry, In certain parameter regimes, this modification is sufficient to remove the classical singularity at the quantum level, according to the HM criterion. The resolution is, nonetheless, non-generic: quantum regularity is achieved only for specific relations between the mass parameter  $m$ and the YM charge $Q$, for which the effective potential governing wave propagation render the spatial operator essentially self-adjoint. \\

For higher dimensions $D \geq 6$, the situation is markedly different. Although the outer singularity can be healed quantum mechanically for particular values of the mass parameter $m$, the central singularity generally remains quantum mechanically singular. Thus, in these higher - dimensional YM space-times, singularity resolution via the HM mechanism is highly constrained and is not realized in the generic case. \\

A natural extension of this work is to explore the robustness of the HM conclusions by employing different quantum probes. The present study focuses on scalar (bosonic wave, spin-0 ) wave propagation: however, fermionic and higher-spin fields may interact with the near-singularity geometry in qualitatively different ways. Since essential self-adjointness of the underlying Hamiltonian can depend sensitively on the structure of the probe field, such analysis could reveal whether the observed quantum singularity is an intrinsic feature of the space-time or merely a property of the specific field considered. Investigating these alternative probes may therefore provide a more complete understanding of quantum singularity resolution in Yang-Mills coupled gravity systems.

\textbf{Appendix A. Kretschmann scalars }\\

In this appendix, we provide the explicit forms of the Kretschmann scalars corresponding to the three different metrics for all dimensional casses examined in this study. The following equations  the Kretschmann scalars corresponding to the Einstein-Yang-Mills, Einstein-Maxwell-Yang-Mills, and Einstein-Gauss-Bonnet-Yang-Mills solutions, respectively.

\begin{equation}
\hspace*{-10mm} \mathcal{K}_{EYM}=
\begin{cases}
\frac{2}{r^4}
\begin{aligned}[t]
&\Bigg(
\frac{ 6\left(m+2 Q^2 \ln (r)\right)^2}{r^4}
+ \frac{12\left(m+2 Q^2 \ln (r)-Q^2\right)^2}{r^4} \\
&- 3 m - 6 Q^2 \ln (r) + 5 Q^2
\Bigg)
\end{aligned}
\\[5pt]
\hfill \text{if } D=5 \hfill \\[15pt]
\frac{(D-3)}{(D-5)^2r^{2 (D+4)}}
\begin{aligned}[t]
&\Bigg(
2 (D-5)^2 (D-2)^2 m^2 r^{10} \\
&+ (D-5) (D-2) m r^{D+5} \big(12 (D-3) Q^2-(D-5) r^4\big) \\
&+ 2 Q^2 r^{2 D} \big((D-3) (D-2) (D+1) Q^2-3 (D-5) r^4\big)
\Bigg)
\end{aligned}
\\[5pt]
\hfill \text{if } D>5 \hfill
\end{cases}
\end{equation}

\begin{equation}
\hspace*{-28mm} \mathcal{K}_{EMYM}=
\begin{cases}
\frac{2}{9r^{12}}
\begin{aligned}[t]
&\Bigg(
6\left(3 m r^2+6 Q^2 r^2 \log (r)-2 Q^2\right)^2 \\
&+ 12 \left(3 m r^2 - Q^2 (3 r^2 + 4) + 6 Q^2 r^2 \log (r) \right)^2 \\
&-27 m r^8 + 45 Q^2 r^8 - 54 Q^2 r^8 \log (r) + 60 Q^2 r^6
\Bigg)
\end{aligned}
\\[5pt]
\hfill \text{if } D=5 \hfill \\[15pt]
\frac{1}{r^4}
\begin{aligned}[t]
&\Bigg(
\frac{2 (D-3) (D-2)}{r^4}
\left(m r^{5-D} + 2 Q^2 \left(\frac{1}{D-5} - \frac{2 (D-3) r^{8-2D}}{D-2}\right) \right)^2 \\
&+ 2 (D-2) \left(m r^{3-D} + \frac{(D-3) Q^2 \left(\frac{1}{D-5} - \frac{2 r^{8-2D}}{D-2}\right)}{r^2} \right)^2 \\[5pt]
&- (D-2) m r^{5-D} + \frac{4 (D-3) (2D-5) Q^2 r^{8-2D}}{D-2} - \frac{6 Q^2}{D-5}
\Bigg)
\end{aligned}
\\[5pt]
\hfill \text{if } D>5 \hfill
\end{cases}
\end{equation}

\begin{equation}
\hspace*{-28mm} \mathcal{K}_{EGBYM}=
\begin{cases}
\frac{4}{r^{4}\sigma(r)}
\begin{aligned}[t]
&\Bigg(
\pm\frac{4  \alpha  \left(4 \alpha +2 m+4 Q^2 \log (r)-Q^2\right)^2}{r^4 \sigma(r)^{2}}\pm\frac{5}{2}Q^2\mp(\alpha +m)\mp2Q^2 \log (r)\Bigg)\\
&+\frac{1\mp\sigma(r)}{2\alpha }\left(1+\frac{3 \left(1\mp\sigma(r)\right)}{ 2\alpha} \right)  \\
&+ \frac{3 \left(r^4 \left(\sigma(r)\mp1\right)+4 \alpha  Q^2\right)^2}{2\alpha ^2 r^4 \left(8 \alpha  (2 \alpha +m)+16 \alpha  Q^2 \log (r)+r^4\right)}
\end{aligned}
\\[5pt]
\hfill \text{if } D=5 \hfill \\[15pt]
\frac{2}{\Sigma(r)}
\begin{aligned}[t]
&\Bigg(\pm\frac{4 r^2 \bar{\alpha}  \left((D-1) m \bar{\alpha}  r^{-D}+\frac{2 (D-3) Q^2}{(D-5) r^5}\right)^2}{\Sigma^2(r)}\mp\frac{2 (D-3) Q^2}{(D-5) r^4}\pm4 (D-1) m \bar{\alpha}  r^{1-D}\\
&\mp (D-1) D m \bar{\alpha}  r^{1-D}\Bigg)+ \frac{1\mp\Sigma(r)}{\bar{\alpha} }\left(1+\frac{(D-3) (D-2) \left(1\mp\Sigma(r)\right)}{2 \bar{\alpha}} \right) \\[5pt]
&+\frac{2 (D-5) (D-2) r^{4-D} \left(r^D \left(1\mp\Sigma(r)\right)-2 (D-5) m r \bar{\alpha} ^2\right)^2}{8 (D-5) m r^5 \bar{\alpha} ^4+\bar{\alpha} ^2 r^D \left(4 (D-3) Q^2 \bar{\alpha} +(D-5) r^4\right)}
\end{aligned}
\\[5pt]
\hfill \text{if } D>5 \hfill
\end{cases}
\end{equation}

in which $\Sigma(r)=\sqrt{\frac{8 m \bar{\alpha} ^2}{r^{D-1}}+\frac{4 (D-3) Q^2 \bar{\alpha} }{(D-5) r^4}+1}$, $\sigma(r)=\sqrt{\frac{16 \alpha ^2 \left(\frac{m}{2 \alpha }+1\right)}{r^4}+\frac{16 \alpha  Q^2 \log (r)}{r^4}+1}$.

\end{document}